\documentclass[aps,prl,twocolumn,superscriptaddress,showpacs]{revtex4-1}
\usepackage{amsmath}
\usepackage{amsthm}
\usepackage{mathrsfs}
\usepackage{mathtools}
\usepackage{graphicx}
\usepackage{scalerel}
\usepackage{color}
\usepackage{tabularx}
\usepackage{array}
\usepackage{multirow}
\usepackage[percent]{overpic}
\usepackage{yfonts}
\usepackage[normalem]{ulem}
\usepackage{bm}
\usepackage{mathptmx}

\usepackage[colorlinks=true,linkcolor=blue,citecolor=blue,urlcolor=blue]{hyperref}

\newcommand{\dee}{\mathrm{d}}
\newcommand{\rmin}{\mathrm{in}}
\newcommand{\rmout}{\mathrm{out}}

\def\hgzero{\scalerel*{\includegraphics{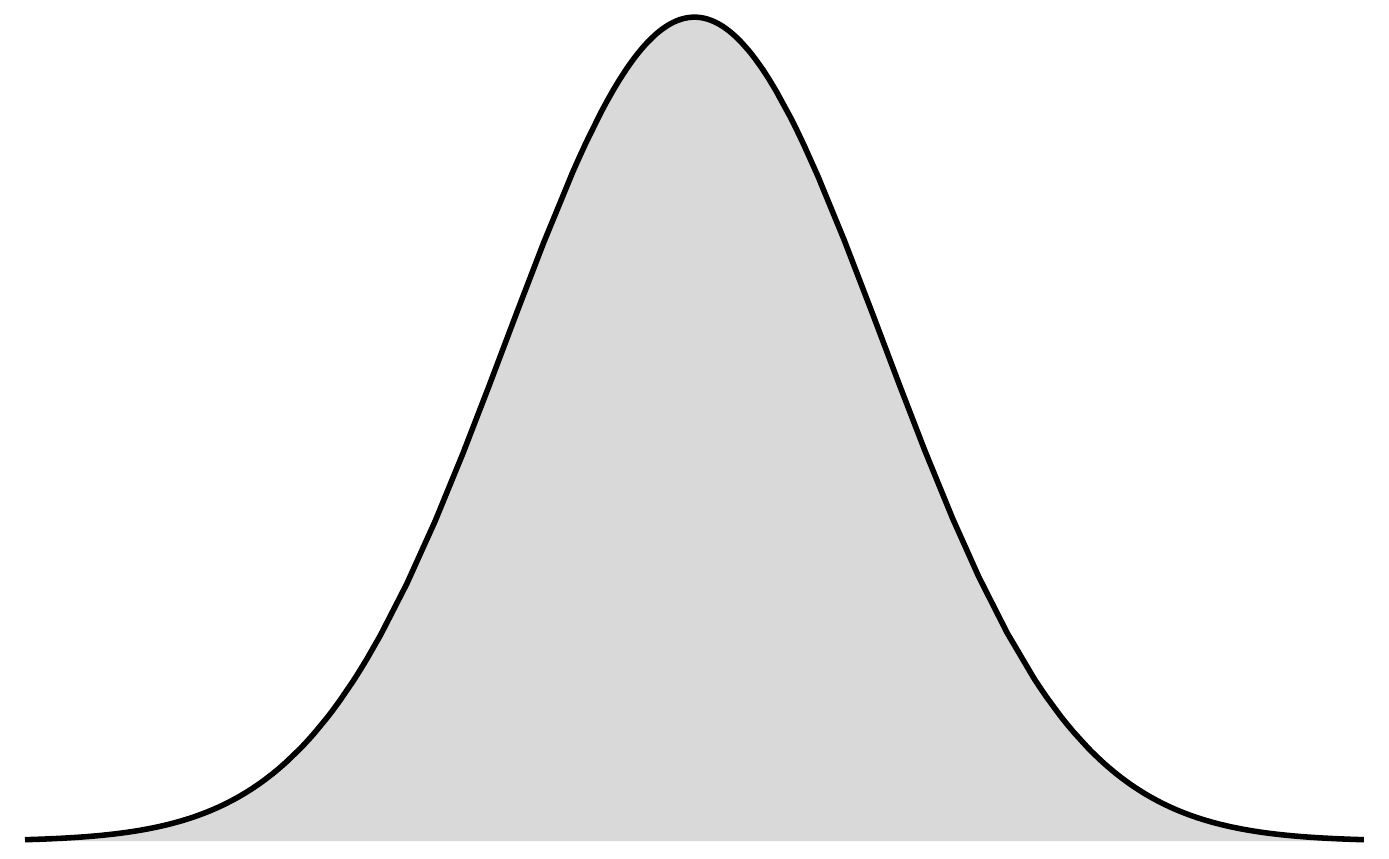}}{p}}
\def\hgone{\scalerel*{\includegraphics{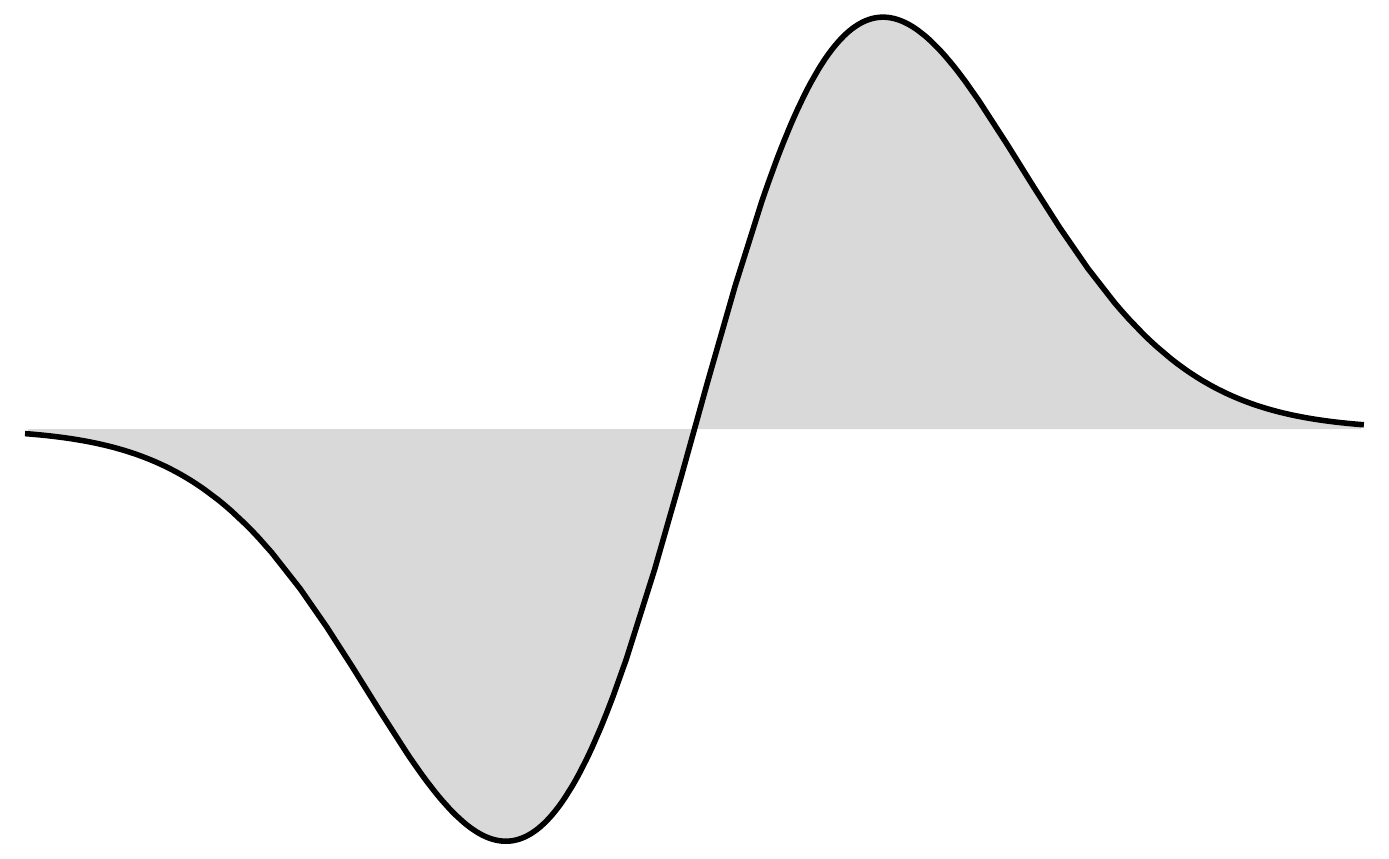}}{b}}

\begin{document}

\title{Quantum-limited time-frequency estimation through mode-selective photon measurement}

\author{J.~M.~Donohue}
\email{john.matthew.donohue@uni-paderborn.de}
\affiliation{Integrated Quantum Optics, Paderborn University, Warburger Strasse 100, 33098 Paderborn, Germany}

\author{V.~Ansari}
\affiliation{Integrated Quantum Optics, Paderborn University, Warburger Strasse 100, 33098 Paderborn, Germany}

\author{J.~\v{R}eh\'a\v{c}ek}
\affiliation{Department of Optics, Palack\'{y} University, 17. listopadu 12, 771 46 Olomouc, Czech Republic}

\author{Z.~Hradil}
\affiliation{Department of Optics, Palack\'{y} University, 17. listopadu 12, 771 46 Olomouc, Czech Republic}

\author{B.~Stoklasa}
\affiliation{Department of Optics, Palack\'{y} University, 17. listopadu 12, 771 46 Olomouc, Czech Republic}

\author{M.~Pa\'ur}
\affiliation{Department of Optics, Palack\'{y} University, 17. listopadu 12, 771 46 Olomouc, Czech Republic}

\author{L.~L. S\'{a}nchez-Soto}
 \affiliation{Departamento de \'{O}ptica, Facultad de F\'{\i}sica, Universidad Complutense, 28040 Madrid, Spain}
 \affiliation{Max-Planck-Institute f\"{u}r die Physik des Lichts, Staudtstrasse 2, 91058 Erlangen, Germany}

 \author{C.~Silberhorn}
 \affiliation{Integrated Quantum Optics, Paderborn University, Warburger Strasse 100, 33098 Paderborn, Germany}

\date{\today}

\begin{abstract}
  By projecting onto complex optical mode profiles, it is possible to estimate arbitrarily small separations between objects with quantum-limited precision, free of uncertainty arising from overlapping intensity profiles. Here we extend these techniques to the time-frequency domain using mode-selective sum-frequency generation with shaped ultrafast pulses. We experimentally resolve temporal and spectral separations between incoherent mixtures of single-photon level signals ten times smaller than their optical bandwidths with a ten-fold improvement in precision over the intensity-only Cram\'er-Rao bound.
\end{abstract}


\maketitle

\emph{Introduction.---} The time-honored Cram\'er-Rao lower bound
(CRLB)~\cite{Cramer:1946aa,Rao:1945aa} is credibly the most
appropriate tool to address the resolution limits for incoherent
imaging, as highlighted in recent years~\cite{Farrell:1966aa,
  Orhaug:1969aa,Helstrom:1970aa,Helstrom:1970ab,Zmuidzinas:2003aa,
  Holmes:2013aa,Motka:2016aa}. This is especially pertinent when photon
shot noise is the dominant noise source (as in, for example,
astronomical observations) and a statistical treatment of resolution
is indispensable. Nonetheless, in spite of these compelling results, Cram\'er-Rao resolution
limits did not demand a great deal of attention until recent
works examined microscopy limitations from a photon-counting perspective~\cite{Bettens:1999aa,Ram:2006aa,Chao:2016aa}.  The chief idea can be formalized
through the Fisher information~\cite{Fisher:1925aa}, which quantifies
the amount of information gained per photon detection and is
directly associated to the CRLB. For direct intensity imaging, the
Fisher information drops to zero for object
separations smaller than the spread of the optical field. This precipitous drop, named Rayleigh's curse, limits the usefulness of photon counting for metrology. 

This line of questioning cleared the way for a fresh reexamination of the problem by Tsang
and coworkers~\cite{Tsang:2016aa,Nair:2016aa,Nair:2016ab,Tsang:2017aa,Tsang:2018aa}. Surprisingly, when one calculates the quantum Fisher
information~\cite{Petz:2011aa} (i.e., optimized over all the possible
quantum measurements), the associated quantum CRLB maintains a fairly
constant value for any separation of the sources. This shows the
potential for parameter estimation of distributions with
precision unaffected by Rayleigh's curse. The key behind these
techniques is phase-sensitive measurement in mode bases other than intensity~\cite{Lupo:2016aa,Rehacek:2017aa}. This has been experimentally demonstrated for spatially separated objects by holographic mode projection~\cite{Paur:2016aa}, heterodyne
detection~\cite{Yang:2016aa,Yang:2017aa}, and parity-sensitive
interferometers~\cite{Tham:2016aa}. As the strong analogy between the space-momentum and time-frequency
descriptions of light has already provided valuable insights and
useful techniques such as temporal imaging~\cite{Bennett:1994aa}, it
is worthwhile to consider the advantages these techniques can offer
when adapted to different domains.

In this Letter, we show that mode-selective measurement can be
harnessed to estimate separations in time and frequency well below the
spread of the source light. In analogy to the Rayleigh limit in space,
this allows us to overcome the Taylor criterion in measuring spectral
separations~\cite{Juvells:2006aa}, which states that the minimum resolvable separation of the
spectral maxima is equal to the half-maximum width. We experimentally realize
this enhancement in both time and frequency estimation settings by
projectively measuring Hermite-Gauss time-frequency modes using
sum-frequency generation with shaped ultrafast pulses in
group-velocity engineered nonlinear waveguides. We explicitly
demonstrate precision below the intensity-only CRLB, establishing
mode-selective measurement as a valuable tool for pushing metrological
limits in multiple physical domains.

Quantum analysis of time-frequency metrological problems has already
provided a plethora of useful tools. In particular, quantum advantages
can be realized in time-of-flight measurement and synchronization by
exploiting entanglement~\cite{Giovannetti:2011aa},
squeezing~\cite{Lamine:2008aa}, and
bunching~\cite{Lyons:2017aa}, and considering quantum
techniques and analysis has inspired classical techniques that
outperform their pre-existing
counterparts~\cite{Kaltenbaek:2008aa}. Additionally, reductions in
the standard quantum limit have been noted using homodyne techniques with
shaped local oscillators in higher-order Hermite-Gaussian
modes~\cite{Jian:2012aa,Thiel:2017aa}. Here, we
show that quantum-inspired metrology finds application in measuring
incoherent source superpositions with either time or frequency
offsets. This form of frequency estimation has natural applications
in, for example, measuring nearly degenerate atomic and stellar
spectral lines, particularly after undergoing inhomogenous
broadening. Precision time measurements find natural applications in
time-of-flight ranging and in probing ultrafast system dynamics.

\emph{Quantum-limited measurements.---}
We formalize the parameter estimation problem under consideration in analogously to the spatial case~\cite{Tsang:2016aa,Paur:2016aa}. Two mutually incoherent (or phase-randomized) light sources with equal intensities emit at optical frequencies $\nu_0\pm\frac{\textgoth{s:}_\nu}{2}$. We assume that the central frequency $\nu_0$ is well-known and that the remaining quantity of interest is the spectral separation, $\textgoth{s:}_\nu$. If the sources have non-negligible spectral bandwidth, the optical spectrum $I(\nu, \textgoth{s:}_\nu)$ of the incoherent mixture as measured on a spectrometer will be \begin{equation}
I(\nu,\textgoth{s:}_\nu)=\frac{1}{2} \left(\left|\psi\left (\nu+ \frac{\textgoth{s:}_\nu}{2} \right )\right|^2+\left|\psi\left(\nu-\frac{\textgoth{s:}_\nu}{2}\right)\right|^2\right),\end{equation} where $\psi(\nu)$ is the spectral amplitude shape. For specificity, we focus on the case of Gaussian spectral amplitudes (frequency-domain point-spread functions) with root-mean-square (RMS) widths $\sigma_\nu$, such that \begin{equation}\psi\left(\nu\pm\frac{\textgoth{s:}_\nu}{2}\right)=\frac{1}{(2\pi\sigma_\nu^2)^\frac{1}{4}} \exp\left[-\frac{\left(\nu-\nu_0\pm\frac{\textgoth{s:}_\nu}{2}\right)^2}{4\sigma_\nu^2}\right].\label{eq:gaussian}\end{equation} 

The standard method of estimating the spectral separation $\textgoth{s:}_\nu$ in the low-luminescence (i.e. photon counting) regime would be to measure the spectral intensity $I(\nu,\textgoth{s:}_\nu)$ on a spectrometer, such as a Fabry-P\'{e}rot interferometer or grating-based spectrograph, and use a fitting or deconvolution algorithm on the integrated photon counts. We quantify the amount of information in-principle available to estimate $\textgoth{s:}_\nu$ with $N$ detected photons (i.e., standard intensity detection) via the Fisher information $\mathcal{F}_\mathrm{std}$, given by \begin{equation}\mathcal{F}_\mathrm{std}= N\int^\infty_{-\infty}\dee\nu\,\frac{1}{I(\nu,\textgoth{s:}_\nu)}\left [ \frac{\partial I(\nu,\textgoth{s:}_\nu)}{\partial \textgoth{s:}_\nu}\right ]^2.\end{equation} The Fisher information quantifies how sensitive the measured quantity $I(\nu,\textgoth{s:}_\nu)$ is to changes in the variable $\textgoth{s:}_\nu$, and can be used to construct the CRLB as $\mathrm{Var}(\hat{\textgoth{s:}}_\nu)\geq1/\mathcal{F}_\mathrm{std}$~\cite{Motka:2016aa}, which defines the absolute minimum mean-squared error (variance) of the estimated separation, $\hat{\textgoth{s:}}_\nu$. For large separations, ${\textgoth{s:}_\nu\gg\sigma_\nu}$, the standard Fisher information is constant, providing a Cram\'{e}r-Rao bounded variance of $\mathrm{Var}(\hat{\textgoth{s:}}_\nu)\geq(4\sigma_\nu^2)/N$. However, when $\textgoth{s:}_\nu\sim\sigma_\nu$, the CRLB bound grows dramatically, diverging as $\textgoth{s:}_\nu/\sigma_\nu$ approaches zero. This behavior is known as Rayleigh's curse in the spatial domain, and is sometimes rephrased as the Taylor criterion in spectral measurements. Note that the exact same ``curse'' applies to estimating incoherent time separations, $\textgoth{s:}_t$, between two pulsed sources through direct timing measurement, for example with autocorrelation or streak-camera techniques~\cite{bradley1971direct}.

The curse can be lifted by performing phase- or parity-sensitive measurements, even though the source fields themselves have no coherent phase relationship. An optimal measurement basis is always provided by the partial derivatives of the amplitude point-spread function~\cite{Rehacek:2017aa}. For the Gaussian point-spread function as in Eq.~\eqref{eq:gaussian}, the optimal measurement is then the Hermite-Gaussian basis~\cite{Tsang:2016aa,Paur:2016aa,Rehacek:2017aa}. For separations $\textgoth{s:}_\nu\lesssim\sigma_\nu$, $\textgoth{s:}_\nu$ can be optimally estimated with only projections onto the first two Hermite-Gauss modes, expressed as \begin{equation}\begin{split} \phi_{\hgzero}(\nu) &= \frac{1}{(2\pi\sigma_\nu^2)^\frac{1}{4}}\exp\left[-\frac{(\nu-\nu_0)^2}{4\sigma_\nu^2}\right] \\ \phi_{\hgone}(\nu) &= \frac{(\nu-\nu_0)}{(2\pi\sigma_\nu^6)^\frac{1}{4}}\exp\left[-\frac{(\nu-\nu_0)^2}{4\sigma_\nu^2}\right] .\end{split}\end{equation} If projective measurements onto these modes can be realized, the estimator $\hat{\textgoth{s:}}_\nu$ has curse-free performance, with $\mathrm{Var}(\hat{\textgoth{s:}}_\nu)\geq(4\sigma_\nu^2)/N$ for arbitrarily small values of $\textgoth{s:}_\nu$. This value agrees exactly with the absolute quantum limit derived from the quantum Fisher information~\cite{Tsang:2016aa}. To include estimation of the centroid, extend the technique to large separations $\textgoth{s:}_\nu\gg\sigma_\nu$, or in cases with unequal-intensity emission, higher-order mode projections may be used~\cite{Rehacek:2017ab}.


\emph{Time-frequency mode selection.---}
The key experimental requirement to enable this advantage is mode-selective projective measurement in the time-frequency domain. We implement such measurements using a technique known as the quantum pulse gate~\cite{eckstein2011quantum, manurkar2016multidimensional, reddy2017engineering, ansari2017temporal, ansari2018tailoring}, a sum-frequency process where a weak input signal is mixed with a spectrally shaped pump pulse to create an upconverted signal in a long nonlinear waveguide. To implement a quantum pulse gate, the input signal and pump pulses must have matched group velocities and the walkoff between the input and upconverted signals must be longer than the length of the input pulses. If these conditions are met, the probability of an upconversion event in the low-efficiency regime given an input spectral amplitude $\psi(\nu)$ and a pump amplitude $\alpha(\nu)$ can be expressed simply as~\cite{eckstein2011quantum,ansari2018tailoring} \begin{equation}P_\alpha\propto\left|\int\dee\nu\,\alpha(-\nu)\psi(\nu)\right|^2.\label{eq:QPGproj}\end{equation} Measuring the upconverted pulse power thereby corresponds to a projective measurement on the broadband time-frequency mode defined by the shape of the pump pulse, $\alpha^*(-\nu)$~\cite{ansari2017temporal}. By counting photons in the upconverted mode while projecting on either the fundamental Gaussian mode or the first-order Hermite-Gaussian mode, we can easily construct an estimator by taking their ratio.


The mode selectivity of the quantum pulse gate is limited by the group-velocity walkoff between the input and upconverted signals, which we define by the walkoff parameter $\Delta=\frac{L}{2}\left(\frac{1}{u_{\rmin}}-\frac{1}{u_{\rmout}}\right)$, where $u_j$ is the group velocity and $L$ is the length of the nonlinear interaction. This walkoff defines the phasematching conditions of the interaction, imposing a RMS bandwidth of the upconverted light of ${\sigma_\mathrm{PM}\approx \frac{0.18}{\Delta}}$. When the input and pump are significantly broader than the phasematching bandwidth and the side lobes arising from the sinc-shaped phasematching curve are filtered out, the ratio of the lowest order Hermite-Gaussian projections is given by \begin{equation}\frac{P_{\hgone}}{P_{\hgzero}}\approx \frac{\sigma_\mathrm{PM}^2}{4\sigma_\nu^2}+\frac{\textgoth{s:}_t^2}{16\sigma_t^2}+\frac{\textgoth{s:}_\nu^2}{16\sigma_\nu^2},\label{eq:QPGestimator}\end{equation} where $\sigma_t$ and $\sigma_\nu$ are the RMS widths of the measurement pulse's temporal and spectral profiles; a derivation of this result is presented in the appendix. If the signal is properly aligned in one of the two degrees of freedom ($\textgoth{s:}_\nu=0$ or $\textgoth{s:}_t=0$) and  $\frac{\textgoth{s:}_\nu}{\sigma_\nu}\gg\frac{\sigma_\mathrm{PM}}{\sigma_\nu}$ or $\frac{\textgoth{s:}_t}{\sigma_t}\gg\frac{\sigma_\mathrm{PM}}{\sigma_\nu}$, Eq.~\eqref{eq:QPGestimator} shows that the square-root of the ratio of projection probability for the first two Hermite-Gauss modes can be used as an exact estimator for the separation between the signals. For separations small enough that the finite phasematching bandwidth cannot be completely neglected, Eq.~\eqref{eq:QPGestimator} can still be inverted to construct a estimator $\hat{\textgoth{s:}}_\nu$ or $\hat{\textgoth{s:}}_t$, although with slightly reduced precision relative to the quantum limit. As the phasematching bandwidth can be much smaller than the input bandwidths~\cite{allgaier2017bandwidth}, the precision of this method can be considerably finer than the broad bandwidth or temporal durations of the pulses being interrogated.
\begin{figure}[t!]
  \begin{center}
      \includegraphics[width=1\columnwidth]{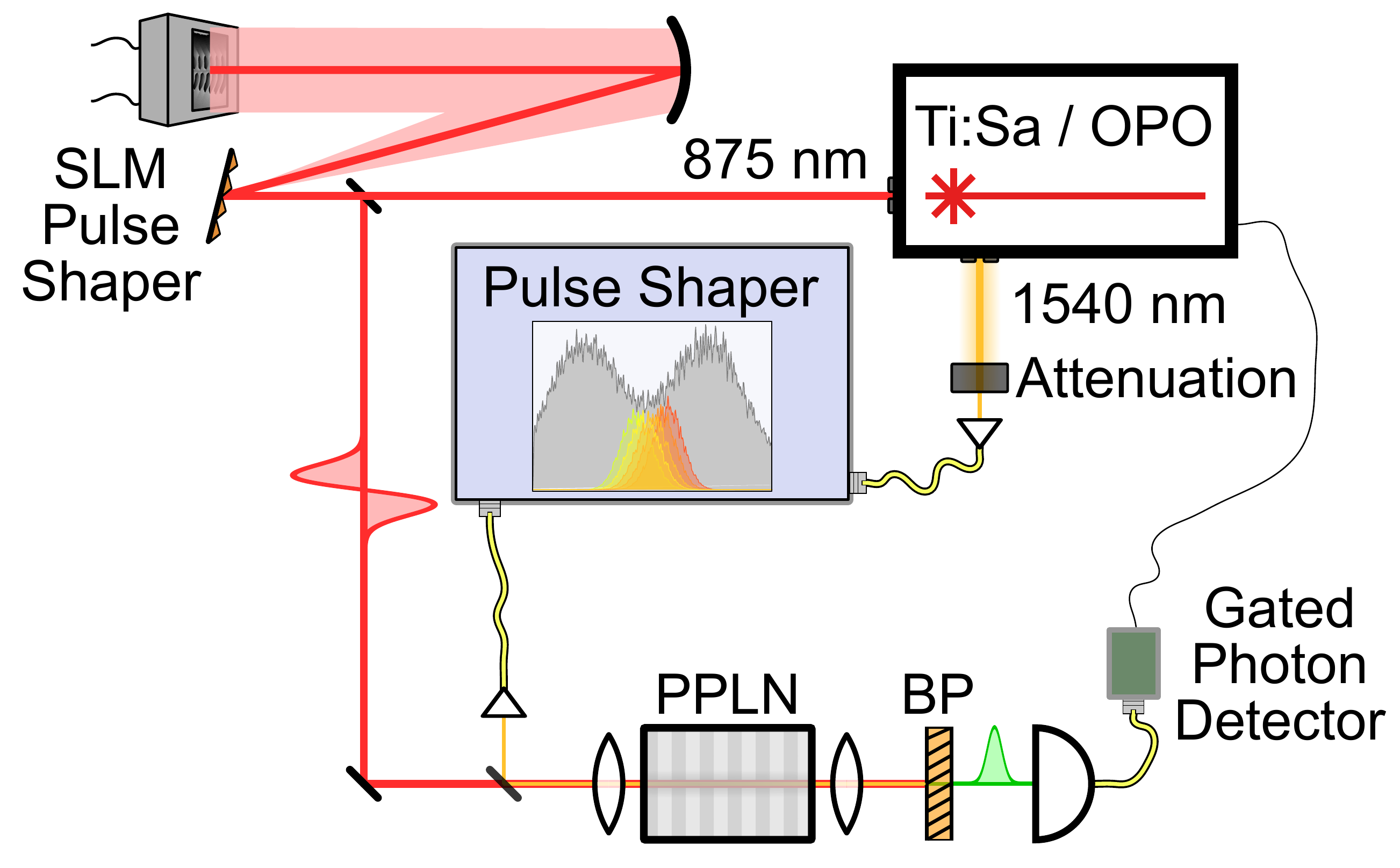}
  \end{center}
 \caption{\textbf{Experimental setup.}  We carve signal pulses with shifting center frequencies and time delays from an attenuated broadband Ti:Sapphire OPO pulse at 1540~nm using a commercial telecommunications pulse shaper. We shape pump pulses at 875~nm into Hermite-Gaussian shapes using a 4f-line with a spatial light modulator (SLM). We then mix the pump and signal pulses in a PPLN waveguide, separate the sum-frequency signal with a 4f bandpass filter (BP), and count photons using an avalanche photodiode gated by a clock pulse from the Ti:Sa.}\label{fig:expsetup}
\end{figure}


\begin{figure}[t!]
\begin{flushleft}
      \hspace{0.15\columnwidth}\begin{overpic}[height=0.8\columnwidth]{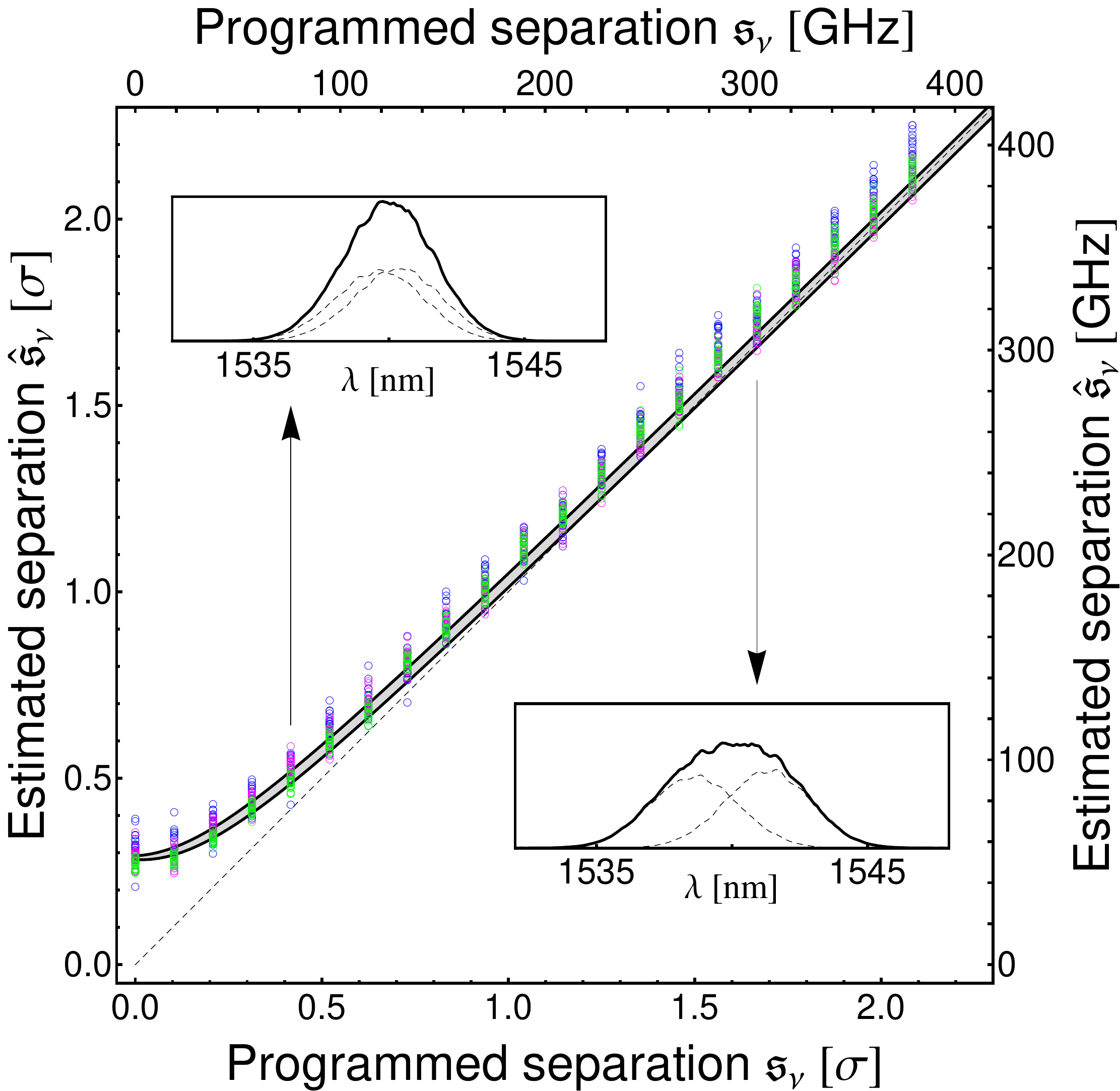}{\put (0,90) {(a)}}\end{overpic}\\
      \vspace{0.1cm}
      \hspace{0.15\columnwidth}\begin{overpic}[height=0.8\columnwidth]{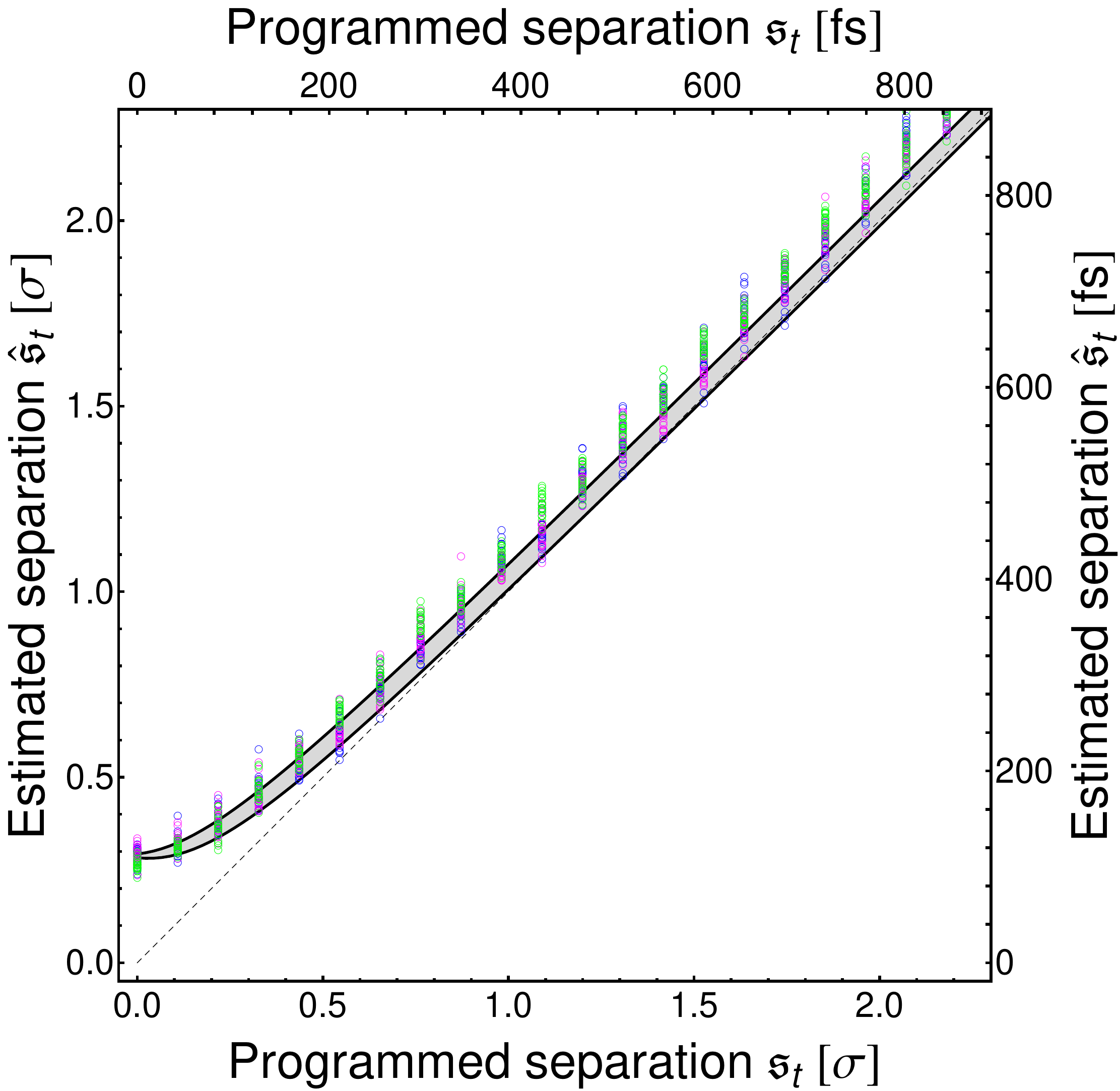}{\put (0,90) {(b)}}\end{overpic}
\end{flushleft}
 \caption{\textbf{Raw estimator from time-frequency mode selection.} The estimator calculated from the measured counts when mixing frequency- or time-shifted pulses is shown above, in (a) and (b) respectively. The solid black lines corresponds to the theoretical expectation given the measured phasematching bandwidth, and the dashed line to the ideal slope-one estimator. The error in the theory curves correspond to instrument setting and bandwidth characterization uncertainty. In both cases, limitations are encountered for separations below 0.2$\sigma$, as expected from the mode-selectivity of the device. The insets on the frequency-measurement plot (a) provide the spectra of the individually shifted signals (dashed) and their incoherent mixture (solid) for programmed separations of $\textgoth{s:}_\nu=0.42\sigma$ and $1.67\sigma$.}\label{fig:rawdata}
\end{figure}

\emph{Experiment.---}  In our experimental apparatus, sketched in Fig.~\ref{fig:expsetup}, we generate shaped input signal and pump pulses from a Ti:Sa laser and optical parameter oscillator (OPO) with a repetition rate of 80~MHz. The strong pump pulses at 875~nm are shaped into Hermite-Gauss modes with a bandwidth of 1.3~nm full-width at half-maximum (FWHM) using a 4f-line with a spatial light modulator (SLM) at the focal plane, with approximately 2~mW coupled into the quantum pulse gate. 

To create frequency- and time-shifted pulses, we carve Gaussian signals with intensity RMS widths of ${\sigma_\nu=182\pm2~\mathrm{GHz}}$ from the approximately $3$-THz FWHM emission of the OPO using a commercial pulse shaper (Finisar 4000S). Frequency shifts are imparted straightforwardly by carving different parts of the OPO spectrum, while time shifts are imparted by programming linear spectral phases with the pulse shaper. The width of the pulses in time was measured to be $\sigma_t={387\pm13~\mathrm{fs}}$ using the quantum pulse gate as an autocorrelator. Neutral density filters were used to attenuate the shaped pulses to approximately 1.1 photons per pulse coupled into the measurement waveguide. The incoherence of the time- and frequency-separated mixtures was assured by switching between positive and negative shifts and mixing the measured results.


To quantum pulse gate was realized by combing the shaped input and pump pulses on a dichroic mirror and coupling them into a 17-mm-long and 7-$\mu$m-wide periodically poled lithium niobate (PPLN) waveguide with a poling period of 4.4~$\mu$m and single-mode propagation at 1540~nm. The spectra of the upconverted light at 558~nm was cleaned with a 4f line to remove phasematching side lobes, resulting in an upconverted bandwidth of $\sigma_{\mathrm{PM}}=28$~GHz, a factor of six smaller than the input light. The internal upconversion efficiency, measured as the depletion of the transmitted signal for the Gaussian projection with zero offset, was approximately 18\%. To reduce background noise due to detector dark counts, the upconverted signal was measured with an avalanche photodiode in coincidence with a clock pulse from the Ti:Sa sampled down by a factor of 50, resulting in an effective experimental repetition rate of 1.6~MHz.


\begin{figure}[t!]
  
\begin{flushleft}
\hspace{0.45\columnwidth}\begin{overpic}[width=0.55\columnwidth]{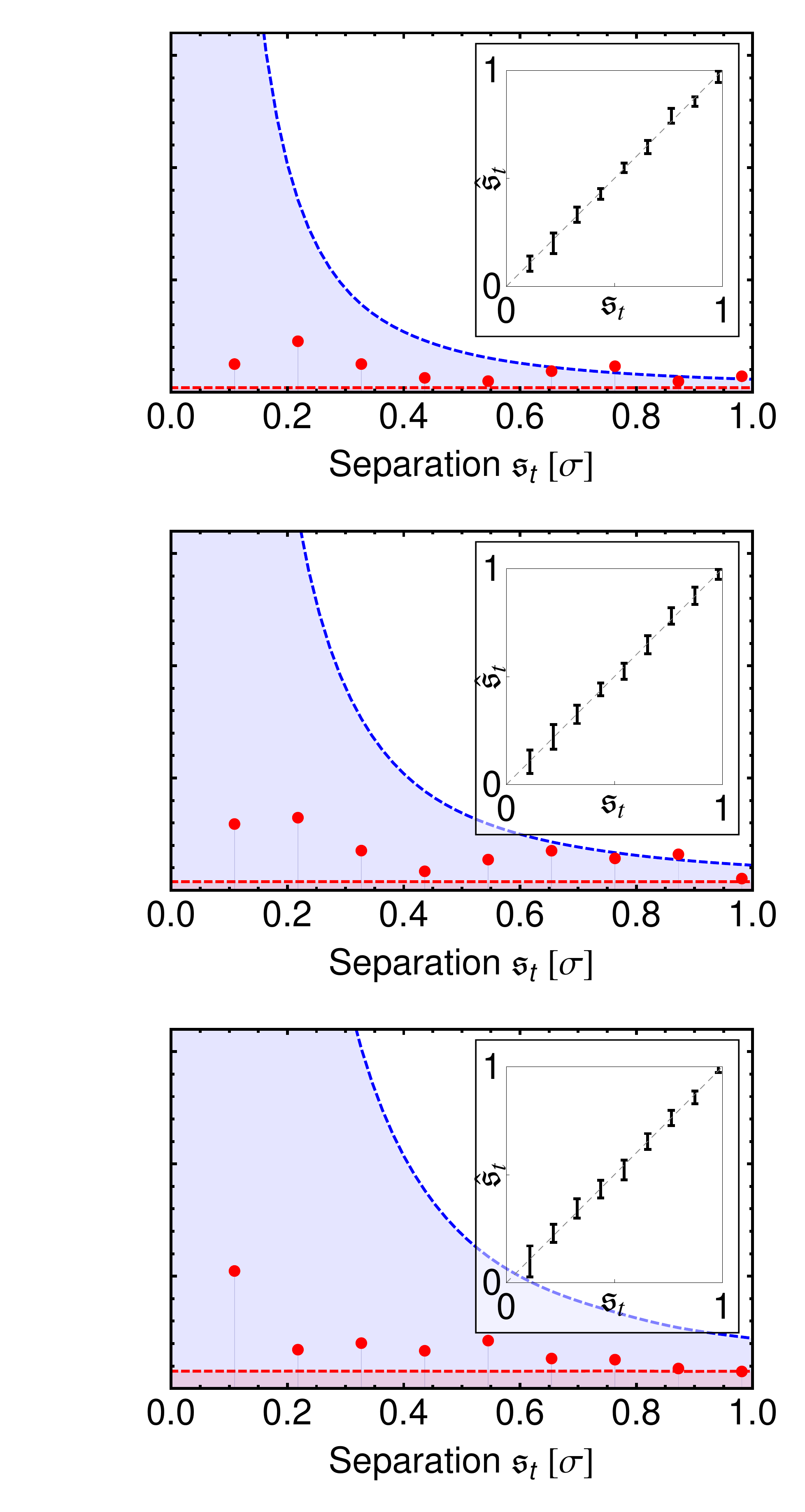}{\put (6.5,98) {(b)}}\end{overpic}\hspace{-1.0\columnwidth}\begin{overpic}[width=0.55\columnwidth]{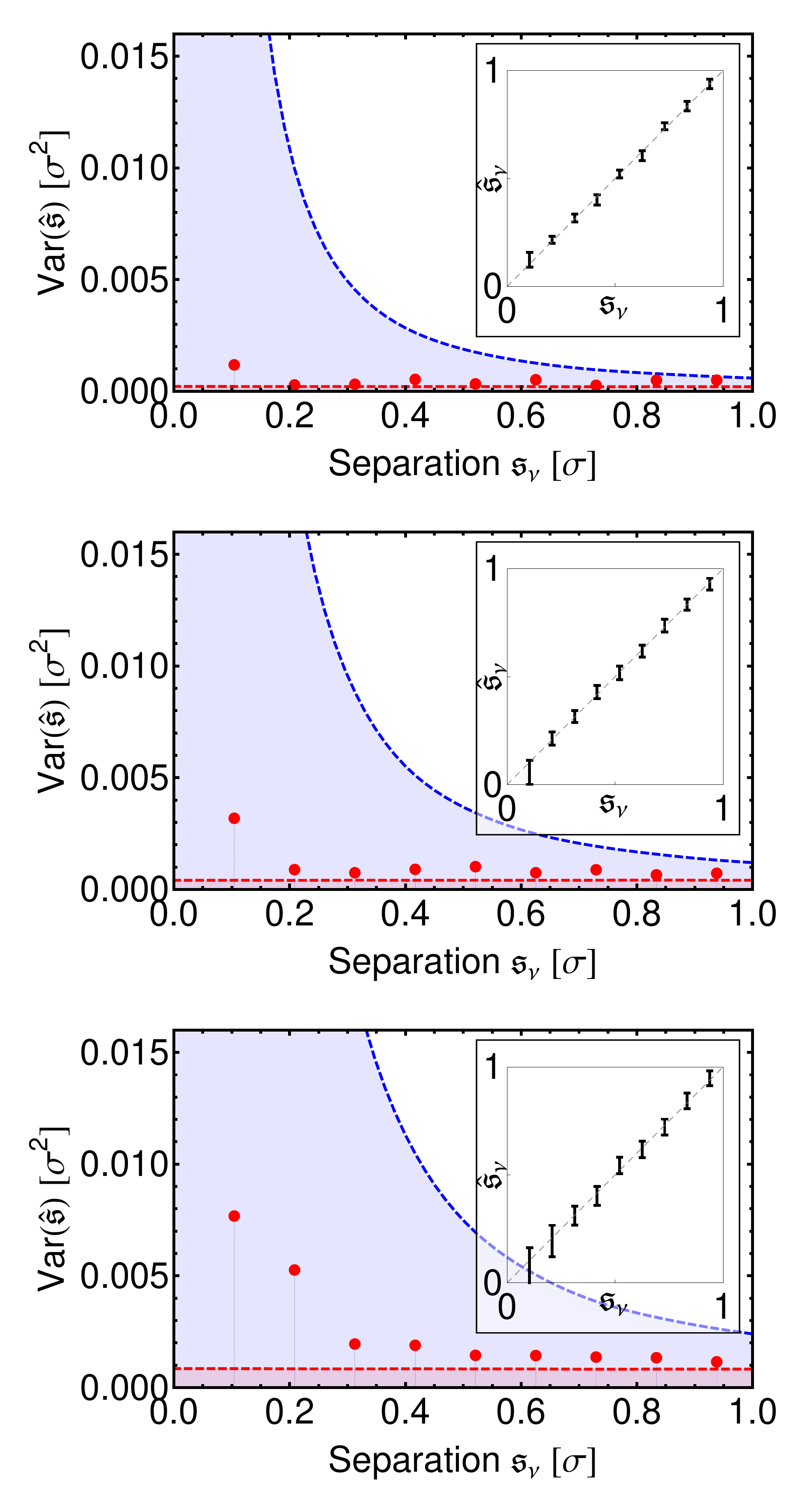}{\put (1,98) {(a)}}\end{overpic}
\end{flushleft}
 \caption{\textbf{Variance of the estimator against the standard bound.} The variance (mean-squared error) of the estimator $\hat{\textgoth{s:}}$ as the (a) frequency separation $\textgoth{s:}_\nu$ or (b) time separation $\textgoth{s:}_t$ is increased. The photon-counting measurements consist of a total of 20,000, 10,000, and 5,000 detection events, from top to bottom. The blue-filled area corresponds to the CRLB for standard intensity detection, and the red dashed line to the quantum limit. Red data points correspond to the variance of the estimator after measurement tomography. The inset shows the estimator after measurement tomography.}\label{fig:qlim}
\end{figure}

\emph{Results and discussion.---}
Twenty separations ranging from $0-2\sigma$ were programmed in both time and frequency during the experiment and each setting was measured 60 times. In addition to controlling the separation, the pulse shaper was also used to attenuate the weak input signal to 100\%, 50\%, or 25\% of its original intensity, to demonstrate the lack of any bias due to background noise. The uncorrected estimator $\hat{\textgoth{s:}}=4\sqrt{P_{\hgone}/P_{\hgzero}}$ from Eq.~\eqref{eq:QPGestimator} is shown in Fig.~\ref{fig:rawdata}. The estimator is seen to reach the expected linear behavior for separations on the same order as the RMS widths, but the imperfect mode selectivity causes small, predictable deviations for very small separations. The observed extinction ratio between the first and zeroth order Hermite-Gauss mode when no separation is found to be ${-10\log_{10}(P_{\hgone}/P_{\hgzero})}={(22.9\pm0.3)}$~dB, corresponding to a minimum estimator value of $\hat{\textgoth{s:}}_{\mathrm{min}}=0.144\pm0.005$.

To construct an unbiased estimator resilient to the imperfect selectivity of our device, we use calibration data from projections onto the first three Hermite-Gauss modes to perform measurement tomography of our technique. Details on the tomography techniques are presented in the appendix. To demonstrate the precision of our technique, in Fig.~\ref{fig:qlim} we show the variance of the calibrated estimator  $\mathrm{Var}(\hat{\textgoth{s:}})$ for both time and frequency measurements while varying the total number of detection events, alongside the standard and quantum CRLBs. The variance is above the quantum limit (in red), owing to mode-selectivity limitations and instabilities. However, it remains below the intensity-only bound $1/\mathcal{F}_{\mathrm{std}}$ for separations well below the point-spread function widths, with an improvement in precision by factor of as high as ten for small separations. 


The above results clearly demonstrate that mode-selective time-frequency measurement can be exploited for precision parameter estimation problems where intensity measurements fail. Notably, the absolute time and frequency scales accessible are not strongly dependent on the scale of the measurement pulses, but rather the material properties, namely the phasematching bandwidth $\sigma_\mathrm{PM}$. In our realization, this corresponded to time and frequency scales of 200~fs and 100~GHz, respectively. The accessible time and frequency scales could be improved either along with the conversion efficiency by increasing the nonlinear interaction length, or at the expense of the detection rate through narrowband filtering of the upconverted signal. Alternative methods based on mode-selective atomic or solid-state Raman memories could provide greater sensitivity, particularly in the frequency domain~\cite{fisher2016frequency,munns2017temporal}. Techniques based on homodyne detection can also provide the necessary mode selectivity in the time-frequency domain~\cite{Lamine:2008aa,polycarpou2012adaptive,roslund2014wavelength}.

We have demonstrated that parameter estimation in the time-frequency domain can benefit greatly from quantum-inspired techniques and analysis. By exploiting time-frequency mode-selective measurement enabled by waveguided nonlinear interactions, we have shown that sub-pulse-width separations can be estimated with precision below the standard CRLB. By adapting these techniques to different scales, this method could find immediate practical use in atomic and stellar spectral characterization and time-of-flight imaging. Future work will explore different mode-selective systems to adapt to specific tangible metrological problems and apply higher-order projections to multi-parameter estimation protocols.

\textit{Acknowledgements} - We thank K.~Bonsma-Fisher, O.~Di~Matteo, J.~Gil-L\'{o}pez, M.~Allgaier, and B.~Brecht for fruitful discussions. This research has received funding from the European Union’s (EU) Horizon 2020 research and innovation program under Grant Agreement No. 665148, the Grant Agency of the Czech Republic (Grant No. 18-04291S), Palack\'{y} University (Grant No. IGA-PrF-2018-003), and the Spanish MINECO (Grant FIS2015-67963-P). J.M.D. gratefully acknowledges support from Natural Sciences and Engineering Resource Council of Canada.

\bibliography{Resolution}

\begin{appendix}

\section{Estimating separations with a quantum pulse gate}\label{sec:QPGestimator}

In this section, we derive Eq.~\eqref{eq:QPGestimator} of the main text, which provides the separation estimators $\hat{\textgoth{s:}}_\nu$ and $\hat{\textgoth{s:}}_t$ for measurements made with a quantum pulse gate. The quantum pulse gate operation relies on a large discrepancy between the group velocities of the input and upconverted signals, which manifests in the energy picture as a much larger input acceptance bandwidth than output signal bandwidth. In practice, this discrepancy is of course finite, which places limitations on the achievable time and frequency resolutions. Here, starting from the basic nonlinear interaction, we outline these limitations.

In our treatment herein, we assume that the three-field interaction takes place inside a single-mode $\chi^{(2)}$ waveguide, such that we may neglect the spatial modes involved. We also assume that we are working in the low-efficiency regime, such that a first-order approach is sufficient~\cite{reddy2017engineering,ansari2018tailoring}. We label the modes as the input (``1''), the QPG pump (``2''), and upconverted output (``3''), and group the central frequencies into the variables as $\tilde\nu=\nu-\nu_0$. In this case, the upconverted spectral amplitude $\gamma(\tilde\nu_3)$ is related to the spectral amplitude of the QPG pump $\alpha(\tilde\nu_2)$ and the input signal $\psi(\tilde\nu_1)$ as \begin{equation}\gamma(\tilde\nu_3)=\theta\int\dee\tilde\nu_1\,H(\tilde\nu_1,\tilde\nu_3)\alpha(\tilde\nu_3-\tilde\nu_1)\psi(\tilde\nu_1),\label{eq:threewavemixing}\end{equation} where energy conservation $\tilde\nu_2=\tilde\nu_3-\tilde\nu_1$ has been accounted for, $\theta$ is a coupling constant representing factors such as the material nonlinearity, and $H(\tilde\nu_1,\tilde\nu_3)$ is the phasematching function, characterized by the relationships between the wavenumbers $k_j(\tilde\nu_j)=\frac{2\pi\nu_jn_j(\tilde\nu_j)}{c}$ of the interacting fields.

If process is phasematched at the central frequencies through periodic poling and chromatic dispersion within each field can be neglected, the phasematching function for an interaction length $L$ can be expressed as \begin{equation}H(\tilde\nu_1,\tilde\nu_3)\propto L\,\mathrm{sinc}\left(\frac{L\left[(k'_1-k'_2)\tilde\nu_1-(k'_3-k'_2)\tilde\nu_3\right]}{2}\right)\end{equation} where $k'_j=\frac{\partial k_j}{\partial\nu_j}\big\rvert_{\nu_{0,j}}=\frac{1}{2\pi u_{j}}$ is inversely proportional to the group velocity $u_{j}$. If the input signal and QPG pump are group-velocity matched $k'_1=k'_2$, the phasematching function simplifies to a function of only the output frequency $\tilde\nu_3$. If we use a bandpass filter to remove the side lobes of the sinc function, we can approximate the phasematching function as a Gaussian, \begin{equation}H(\tilde\nu_3)\approx L\,e^{-\eta\frac{(L(k'_3-k'_1)\tilde\nu_3)^2}{4}}\vcentcolon= L\,e^{-\frac{\tilde\nu_3^2}{4\sigma_{\mathrm{PM}}^2}},\end{equation} where $\sigma_{\mathrm{PM}}$ is the RMS phasematching bandwidth and $\eta\approx0.193$.

We assume that the input signal wavefunction is a Gaussian pulse with some offset $\delta\nu$ from the perfectly phasematched frequencies and a small time delay $\delta t$ relative to the QPG pump pulse, which we express as \begin{equation}\psi(\tilde\nu_1)=\frac{1}{(2\pi\sigma_\nu^2)^\frac{1}{4}}\exp\left[-\frac{(\tilde\nu+\delta\nu)^2}{4\sigma_\nu^2}-i2\pi\tilde\nu\delta t\right].\end{equation} Note that the RMS width of the pulse in time is $\sigma_t=1/(4\pi\sigma_\nu)$. The QPG pump pulse is shaped to the first two Hermite-Gauss temporal modes with bandwidth $\sigma_2$, given by \begin{equation}\begin{split} \alpha_{\hgzero}(\tilde\nu_2) &= \frac{1}{(2\pi\sigma_2^2)^\frac{1}{4}}\exp\left[-\frac{\tilde\nu_2^2}{4\sigma_2^2}\right] \\ \alpha_{\hgone}(\tilde\nu_2) &= \frac{\tilde\nu_2}{(2\pi\sigma_2^6)^\frac{1}{4}}\exp\left[-\frac{\tilde\nu_2^2}{4\sigma_2^2}\right] .\end{split}\end{equation} Substituting these and the phasematching function into Eq.~\eqref{eq:threewavemixing} and finding the relative upconversion probability as $P=\int\dee\nu_3\,|\gamma(\tilde\nu_3)|^2$, the ratio of the upconversion probabilities for the first two modes is found to be \begin{equation}\begin{array}{ccl} \frac{P_{\hgone}}{P_{\hgzero}}&=& \sigma_2^2\left[\frac{\sigma_\nu^2+16\pi^2\delta t^2\sigma_\nu^2+\sigma_2^2}{(\sigma_\nu^2+\sigma_2^2)^2}+\frac{\delta\nu^2-\sigma_\nu^2-\sigma_2^2-\sigma_{\mathrm{PM}}^2}{(\sigma_\nu^2+\sigma_2^2+\sigma_{\mathrm{PM}}^2)^2}\right]\\ &\overset{\sigma_2=\sigma_\nu}{=}& \frac{\sigma_\mathrm{PM}^2}{2(2\sigma_\nu^2+\sigma_\mathrm{PM}^2)} + 4\pi^2\delta t^2\sigma_\nu^2+\frac{\delta\nu^2\sigma_\nu^2}{2(2\sigma_\nu^2+\sigma_\mathrm{PM}^2)^2}\\ &\overset{\sigma_\nu^2\gg\sigma_\mathrm{PM}^2}{\approx}&\frac{\sigma_\mathrm{PM}^2}{4\sigma_\nu^2} + \frac{\delta t^2}{4\sigma_t^2} + \frac{\delta\nu^2}{4\sigma_\nu^2}.\label{eq:QPGestimatorDerivation}\end{array}\end{equation} To get from the first line to the second, we have set the bandwidth of the QPG pump to be equal to the input signal, ensuring that the two pulses have matched temporal-mode bases. To get from the second line to the third, we have assumed that the phasematching bandwidth is narrower than the input pulses, such that ${2\sigma_\nu^2+\sigma_\mathrm{PM}^2\approx2\sigma_\nu^2}$. Since $P_{\hgzero}$ and $P_{\hgone}$ are both symmetric functions of $\delta\nu$ or $\delta t$, Eq.~\eqref{eq:QPGestimatorDerivation} holds for incoherent mixtures of positive and negative shifts, and Eq.~\eqref{eq:QPGestimator} can be retrieved by substituting $\delta\nu\mapsto\frac{\textgoth{s:}_\nu}{2}$ and $\delta t\mapsto \frac{\textgoth{s:}_t}{2}$. It is apparent that the minimum resolvable shift will be on the order of $\sigma_\mathrm{PM}$ in frequency and $\frac{\sigma_\mathrm{PM}}{\sigma_\nu}\sigma_t$ in time, and that any misalignment in frequency or time will adversely effect the resolution of measurements in the other setting.

\section{Measurement tomography methods}\label{sec:MeasTomo}

In this section, we describe the measurement tomography method used to retrieve an accurate separation estimator from the directly measured data. To characterize the device, we implement projections onto the first three Hermite-Gauss modes, where ``ideal measurement'' can be described by projections of the input signal on the three lowest-order Hermite-Gauss modes HG$_0$, HG$_1$, and HG$_2$. We denote $q_j(\textgoth{s:})$ the probability of the $j$th measurement output given the true separation is $\textgoth{s:}$.  For a Gaussian point-spread function (PSF) of width $\sigma$, this probability reads
\begin{equation}
\label{basis}
q_j(\textgoth{s:})=\frac{1}{k!}\left(\frac{\textgoth{s:}}{4\sigma}\right)^{2j} e^{-\left(\frac{\textgoth{s:}}{4\sigma}\right)^2},\qquad j=0,1,2\,.
\end{equation}
Due to unavoidable imperfections, the actual detection probabilities $p_j(\textgoth{s:})$ differ slightly from  $q_j(\textgoth{s:})$ and the measurement device needs to be characterized before using. Assuming the setup works well, that is, the differences between the actual and target distributions $p_j(\textgoth{s:})$ and $q_j(\textgoth{s:})$ are small, we expand the former using the latter as a basis as follows
\begin{equation}
p_j(\textgoth{s:})=\sum\limits_{k=0}^M c_{jk}\, q_k(\textgoth{s:}),\qquad j=0,1,2\,.
\end{equation}
Having repeatedly measured a set of known separations $\textgoth{s:}=\{\textgoth{s:}_1,\textgoth{s:}_2,\ldots,\textgoth{s:}_N\}$, the probabilities $p_j$ can be estimated by the corresponding relative frequencies $f_j=\langle n_j\rangle/\sum_j \langle n_j\rangle$. Denoting further $f^j_\alpha= f_j(\textgoth{s:}_\alpha)$, $q_{k\alpha}= q_k(\textgoth{s:}_\alpha)$, and $c^j_k=c_{jk}$,
we obtain three sets of linear equations to be solved for the set of unknown detection coefficients $c^j_k$
\begin{equation}
f^j_k=\sum_k q_{k\alpha} c_k^j,\qquad j=0,1,2\,.
\end{equation}
The pseudo-inverse can be used to obtain standard solutions minimizing the $L_2$ norm,
\begin{equation}
\mathbf{c}^j=Q^{+} \mathbf{f}^j,\qquad j=0,1,2\,.
\end{equation}
It turns out just a few ($M\approx 4$) basis functions in Eq.~\eqref{basis} are required to observe excellent fits of the detected relative frequencies $f_j$ in terms of the corresponding theoretical models $p_j$ for all measured separations in the region of interest $\textgoth{s:}\in [0,2]$.

We next proceed to the parameter estimation step using our characterized measurement. Each measurement returns a three numbers, $n_0$, $n_1$, and $n_2$. Assuming Poissonian statistics, the separation is estimated by maximizing the log-likelihood
\begin{equation}
\hat{\textgoth{s:}}=\arg\max\limits_{\textgoth{s:}}\left\{ \sum_j n_j \log\left[\frac{p_j(\textgoth{s:})}{\sum_{j'} p_{j'}(\textgoth{s:})}\right]\right\}\,
\end{equation}
subject to $\hat{\textgoth{s:}}\ge 0$ using a suitable optimization tool. Finally, for every true separation we calculate the statistics of the estimates and compare the measurement errors to the relevant classical and quantum resolution limits.

\end{appendix}
\end{document}